\documentclass[
 reprint,
 amsmath,amssymb,
 aps,
]{revtex4-2}

\usepackage{graphicx}
\usepackage{dcolumn}
\usepackage{xcolor}
\usepackage{bm}
\usepackage{float}

\begin{document}

\preprint{APS/123-QED}

\title{
Phase behavior of solvent-nematogen mixtures
}

\author{Sullivan Bailey-Darland}
\author{Takumi Matsuzawa}
\author{Eric R. Dufresne}
\affiliation{%
 Department of Physics, Cornell University
}

\date{\today}

\begin{abstract}
Liquid mixtures with a nematogen can undergo both fluid phase separation and a transition from an isotropic to a nematic state.
These phase transitions can couple and lead to phase behavior distinct from simple liquid mixtures or pure liquid crystals.
We  measured the phase behavior of mixtures of  a nematogen (5CB) with simple liquid solvents (squalane and/or squalene). 
We observed two distinct kinds of binary phase diagrams:  with and  without  a region of isotropic-isotropic coexistence.
Varying the ratio of squalene to squalane,  we continuously tuned the
phase boundaries of the apparent binary system and revealed a region of three-phase coexistence.
A  mean-field model  combining classical models of liquid mixing and nematic ordering  quantitatively describes both binary and ternary phase behavior.
This simple model predicts a range of topologically complex ternary phase diagrams and extends naturally to systems with more components.
\end{abstract}

\maketitle

\section{Introduction}

Fluid phase separation plays an important role in many applications, from materials fabrication \cite{werber2016materials, vauthier2009methods, fernandez2021putting} to liquid purification \cite{gupta1996liquid, vander2013homogeneous}. 
In recent years, research has focused on the emerging role of fluid phase separation in cell biology \cite{banani2017biomolecular, lasker2022material, gouveia2022capillary, martin2020valence}. 
Active research is extending classic models of liquid-liquid phase separation to account for new physics, including  non-equilibrium effects due to biochemical activity \cite{bauermann_theory_2025,zwicker2022intertwined}, the high dimensionality of phase space due to the presence of many components \cite{jacobs2017phase,mao2019phase,zwicker2022evolved,matsuzawa2026metabolites}, and elasticity \cite{tanaka1997viscoelastic,style2018liquid,rosowski2020elastic, ronceray2022liquid,fernandez2024elastic,qiang2024nonlocal,fernandez2026thermodynamics}.
Another fascinating extension of fluid phase separation arises when one phase can undergo an additional phase transition.
This offers the potential for alternate mechanisms of demixing and new ways to design phase diagrams.

Compositional phase transitions are coupled to other phase transitions in a range of systems.
Certain metallic systems can demix while molten or phase-separate into a solid and liquid phase, depending on the composition and temperature \cite{ratke_liquid_1995}. 
Colloidal systems can have ``liquid-gas'' coexistence interrupted by crystallization \cite{lekkerkerker1992phase, wolde_enhancement_1997}.
A mixture of He$^3$ and He$^4$  phase-separates into a  He$^4$-rich superfluid and a He$^3$-rich liquid at low temperatures \cite{griffiths_thermodynamics_1970}.
Lipid membranes can both demix and undergo a transition between the liquid ordered and liquid disordered states \cite{garbes_putzel_phenomenological_2008, wolff_thermodynamic_2011}.
Crystallization and phase-separation can compete in polymer-containing mixtures \cite{li_competition_1991, inaba_morphology_1986, levitsky_bridging_2021, siber_interplay_2026}.
In all these systems, compositional differences can be driven by structural phase transition  from a liquid to a new state (\emph{e.g.} superfluid, crystal, or liquid crystal.).

The coupling of nematic ordering to liquid-liquid phase separation is a fascinating example  of this broad phenomenon.
Consider mixtures of a \emph{nematogen}, a molecule/particle that can undergo an alignment transition, with simple  solvent molecules. 
In a pure  nematogen, the molecules align below a threshold temperature, forming a \emph{nematic} liquid phase.
Dilution of the nematogen with a solvent suppresses alignment (reduces the temperature of the isotropic to nematic transition) \cite{gennes_physics_1975}. 
Alignment of the nematogen can reject solvent and lead to multi-phase coexistence.

The coupling between phase separation and alignment can have significant effects on the structure of the phase diagram.
Binary mixtures of two simple liquids (solvents) results in two co-existing isotropic phases.
The co-existing phases are simply distinguished by their composition.
When a nematogen is added to the mixture, the phases can differ in both composition and orientation. 
Early work on nematogen-solvent mixtures focused on the effect of impurities on thermodynamic properties of the nematic transition \cite{anisimov_anomalies_1980} and two-phase coexistence near the nematic transition \cite{dave_mixed_1954}.
Recent experiments have demonstrated the variety of phase coexistences possible in binary nematogen-solvent mixtures \cite{reyes_isotropicisotropic_2019, serrano_phase_2018} and on fluctuations near  phase boundaries \cite{shimada_phase_2021, kalabinski_phase_2023}.
Applications for the controlled formation of nematic droplets \cite{patel_long_2023, patel_temperature-induced_2021} and liquid-liquid extraction \cite{serrano_phase_2018} have also been investigated.
More generally, mixtures with liquid crystalline properties have been used to make field-controlled emulsions \cite{roh_biphasic_2025}, nematic droplets with complex morphologies \cite{weirich2017liquid,wei_molecular_2019, peddireddy_self-shaping_2021}, and networks of smectic filaments \cite{morimitsu_spontaneous_2024, browne_structural_2025}.

In this paper, we describe experiments and theory broadly exploring the phase behavior of nematogen-solvent mixtures. Using polarized-optical microscopy, we measure a pair of model binary nematogen-solvent mixtures (5CB-squalane and 5CB-squalene) and observe two qualitatively different phase diagrams. 
By blending the solvents in varying ratios, we can continuously tune the structure of phase diagrams.
In this ternary system,  three phases (nematic-isotropic-isotropic) can stably coexist.
To explain these results, we develop a mean-field model for arbitrarily complex nematogen-solvent mixtures. 
This model is able to quantitatively reproduce our measurements with suitably chosen parameters.
Further, it predicts a range of experimentally accessible phase diagrams with complex topologies made from a nematogen and a two solvents.

\section{Results}

\subsection{Binary mixtures of a liquid crystal and solvent}\label{sec:binary_expt}

\begin{figure*}[t]
\includegraphics[width=0.8\textwidth]{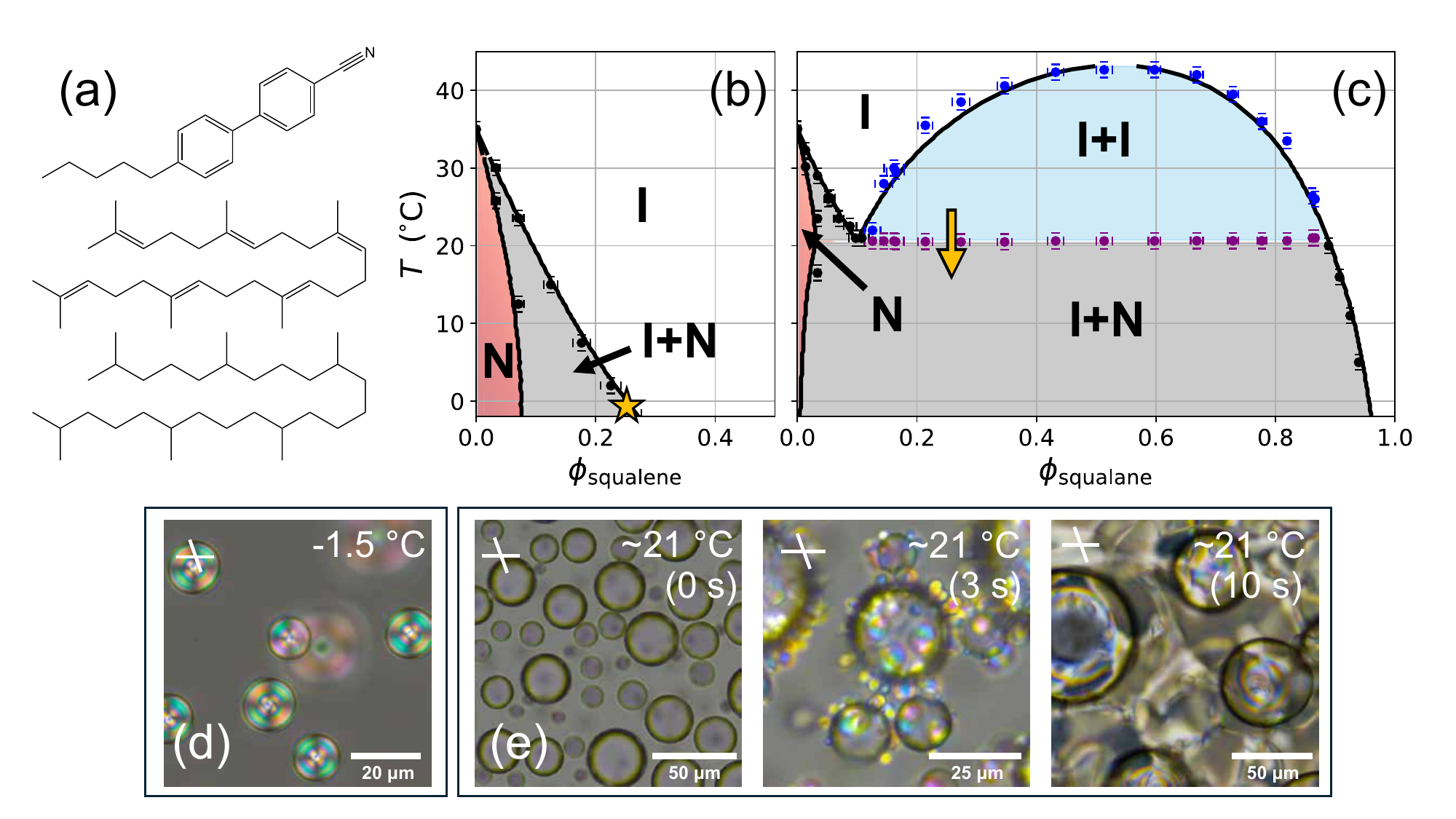}
\caption{\label{fig:binary_data} \textbf{
Binary mixtures of a nematogen and solvent exhibit two kinds of phase diagrams.}
(a) Molecular structures for 5CB (top), squalene (middle), and squalane (bottom).
(b, c) Phase diagrams for binary mixtures of 5CB-squalene and 5CB-squalane, respectively. 
The  shading indicates a nematic liquid (red), isotropic liquid (white), isotropic-nematic coexistence (gray), or isotropic-isotropic coexistence (blue).
Black curves in panels (a) and (b) are fits to theory, described in section \ref{sec:model}.
(d) Image of nematic-isotropic coexistence for $\phi_{\mathrm{squalene}}=0.25$ at -1.5$^\circ$ C with partially crossed polarizers, indicated with a star on panel (a).
(e) Images of the mixture with $\phi_{\mathrm{squalane}}=0.25$ at $\approx21^\circ$ C while cooled at $2^\circ$ C/min, going from isotropic-isotropic to isotropic-nematic coexistence. 
The cooling direction is indicated with an arrow on panel (b).
}
\end{figure*}

From a review of previous measurements, we found  two classes of phase diagrams for mixtures of nematogen and solvent.
The isotropic state  either remains homogeneous  \cite{ seregin_phase_2020, dave_mixed_1954} or  demixes into two isotropic liquids \cite{reyes_isotropicisotropic_2019, serrano_phase_2018, kalabinski_phase_2023, anisimov_anomalies_1980}.
We label these as mixtures of nematogens with ``good" and ``bad" solvents, respectively.
With an eye toward systematically tuning the phase behavior through solvent quality,  we identified experimentally convenient sets of a nematogen, good solvent, and bad solvent.
5CB (4’-pentyl-4-cyanobiphenyl) is a natural choice for the nematogen, as it is easily available, widely studied, and has a nematic transition near room temperature \cite{goodby20244}.
Its structure is shown in Fig. \ref{fig:binary_data}a (top).
Squalene and squalane are convenient choices for solvents \cite{roh_biphasic_2025, morimitsu_spontaneous_2024, browne_structural_2025}. 
Both of these branched alkanes are non-volatile and readily available. 
Squalene and squalane are nearly identical, with varying degrees of saturation, as seen in Fig. \ref{fig:binary_data}a.
Despite these small differences, we will see that have very different compatibilities with 5CB. 

We measured the phase boundaries of these nematogen-solvent mixtures  using temperature-controlled polarized-optical microscopy, as described in the methods.
Squalene is a ``good solvent" for 5CB.  As shown in in Fig. \ref{fig:binary_data}b,
it has a single region of demixing which extends from the nematic transition of pure 5CB at $\approx35^\circ$ C  where the mixture phase separates into isotropic and nematic phases.
In all these measurements, cooling into this region results in the nucleation and growth of nematic droplets.
An image of the resulting droplets are shown in Fig. \ref{fig:binary_data}d, with partially crossed polarizers to show the nematic character of the droplets. 
Squalane is a ``bad solvent'' for 5CB. 
As shown in  Fig \ref{fig:binary_data}c, the mixture can phase separate into two isotropic phases at moderate  solvent concentrations. 
At low solvent concentrations, 
a region of isotropic-nematic coexistence extends from the nematic transition of pure 5CB, similar to the LC-good solvent phase diagram. 
These two regions intersect at $\approx 21^\circ$ C, below which there is coexistence between isotropic and nematic phases.
All two-phase isotropic-isotropic mixtures become isotropic-nematic mixtures at this temperature.

 The region of isotropic-nematic coexistence extending from the nematic transition of the pure nematogen, which we observed in both systems, is a well-known and generic feature of the dilution of a nematogen with a solvent \cite{gennes_physics_1975, anisimov_critical_1991}.  
As dilution increases, the nematic phase appears at a lower temperatures and the temperature range of two-phase coexistence increases.
The difference in composition between the isotropic and nematic phases varies with the choice of solvent.

The mixture of a nematogen with a ``bad solvent" enables three-phase coexistence.
The two regions of demixing in the 5CB-squalane mixture intersect at $\approx 21^\circ$ C.
This is a triple point, where all three phases  exist in equilibrium.
For a system with only two components, this coexistence should occur at a single temperature, which makes it hard to observe in equilibrium. 
However, it is easy to observe dynamically as the mixture is cooled slowly through the triple point.
This is demonstrated for a sample cooled from isotropic-isotropic coexistence through the triple point, shown in Fig. \ref{fig:binary_data}e.  
Once the sample reaches the triple-point temperature, nematic droplets nucleate within one of the two isotropic phases and rapidly grow to occupy the entire phase.
This generally happens in a matter of seconds.

\subsection{Ternary mixtures of a nematogen and two solvents}\label{sec:ternary_expt}

Previous work on the phase behavior of nematogen solvent mixtures were distinguished by discrete changes to their molecular structure
\cite{anisimov_anomalies_1980, seregin_phase_2020, dave_mixed_1954, reyes_isotropicisotropic_2019, serrano_phase_2018, kalabinski_phase_2023}.
In this section, we demonstrate that one can continuously tune the onset of phase separation and the isotropic-nematic transition using solvent mixtures.

\begin{figure}[t]
\includegraphics[width=\columnwidth]{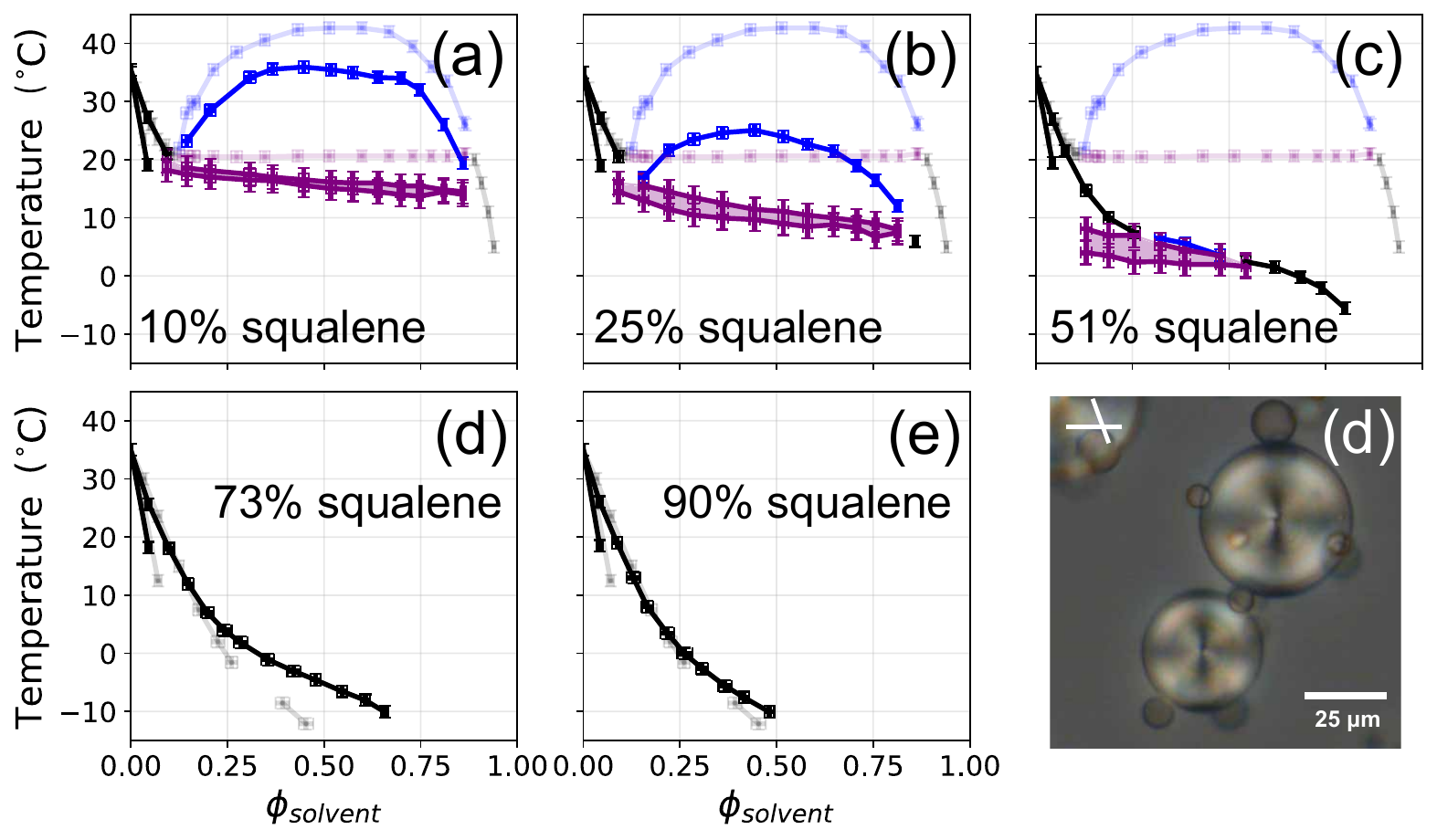}
\caption{\label{fig:ternary_expt} \textbf{Phase behavior of a ternary mixture of nematogen and solvents.} 
(a-e) Phase diagrams for 5CB and squalane-squalene solvent mixtures, with solvent composition varying from 10\% (a) to 90\% (e) squalene, by volume. 
Data points indicate the phase boundaries; black for the isotropic-nematic region, blue for the isotropic-isotropic region, and purple for the three-phase isotropic-isotropic-nematic region.
Curves connect data points to guide the eye.
The phase diagrams of 5CB-squalane and 5CB-squalene are plotted transparently for reference in the top and bottom rows, respectively.
(d) Image taken with partially-crossed polarizers of three-phase coexistence for $\phi_{solvent}=0.25$ in the 51\% squalene mixture at $\approx 7^\circ$ C while cooling at $0.5^\circ$ C/min.
}
\end{figure}

We work with the three molecules explored in the previous section: the nematogen 5CB, the ``good solvent'' squalene and the ``bad solvent'' squalane. 
Squalene and squalane are highly miscible.
Pseudo-binary phase diagrams of this ternary system (temperature versus volume fraction of solvent mixture) are shown in Fig. \ref{fig:ternary_expt}.
As anticipated, binodal curves vary smoothly between the two binary phase diagrams, shown in Fig. \ref{fig:ternary_expt}a-e.
The squalane-rich mixtures are relatively ``bad solvents'' (Fig. \ref{fig:ternary_expt}a,b,c), with the isotropic-isotropic binodal indicated by the blue data points. 
As the solvent quality improves (increasing squalene content),
the isotropic coexistence region shrinks.
These data suggest that solvent blends could been treated as an ``effective solvent'', with a solvent interaction parameter between the two  pure solvents.

However, this simplistic view breaks down with deeper quenches, where we find regions of stable three-phase coexistence for solvent mixtures that have both isotropic-isotropic and isotropic-nematic coexistence (shown in purple in Fig. \ref{fig:ternary_expt}a,b,c).
Error bars are higher for the boundaries of the three-phase region  because it is more difficult to determine the phase transition when the samples are already phase separated.
A representative image of three-phase coexistence is shown  in Fig. \ref{fig:ternary_expt}d, with
 isotropic droplets attached to the surface of a nematic droplet.
The three-phases are stable at fixed temperature (see Appendix Fig. \ref{fig:appendix_three_phase_overnight}). 

\subsection{Mean field model of multi-component mixtures including nematogens}\label{sec:model}

To make sense of our experimental results, and provide a general framework for designing  nematogen-solvent mixtures, we developed a theoretical model combining demixing and alignment.
We combine two canonical mean-field models: multi-component Flory-Huggins theory of phase separation \cite{mao2019phase} and the Maier-Saupe theory \cite{maier1959einfache,maier1960einfache} of the nematic transition. 
This approach results in a free energy model that closely resembles the model in \cite{brochard_phase_1984} for binary nematogen mixtures.

The isotropic part of the free energy depends only on the composition and temperature of the mixture. The composition of the mixture is described as a set of volume fractions $\{\phi_i\}=\{\phi_1, \phi_2, ..., \phi_N\}$, for $N$ different species. The contribution to the free energy density derived from the Flory-Huggins model is
\begin{multline}\label{eq:FH}
g_{\mathrm{FH}}=\sum_i\frac{k_{\mathrm{B}}T}{v_i}\phi_i\ln\phi_i \\ + \sum_{i< j}\phi_i\phi_j\left(A_{ij} + B_{ij}\,T\right) + \sum_{i< j \le k}C_{ijk}\phi_i\phi_j\phi_k
\end{multline}
The first term describes the entropic benefit of mixing, and contains the molecular volumes $v_i$.
Phenomenological interaction terms describing the excess free energy from mixing ($A_{ij}$, $B_{ij}$, $C_{ijk}$) appear in the rest of the function, and these determine the shape of the phase boundary.
Basic Flory-Huggins theory includes only a $\chi_{ij}$ term, equivalent to our $A_{ij}$ term.
The $B_{ij}$ and $C_{ij}$ terms are included to describe temperature and concentration dependence of the $\chi$ parameters \cite{erman_critical_1986, orofino_relationship_1957, fernandez2026thermodynamics}.
Positive values imply that the components interact unfavorably.
The equilibrium state of the mixture is calculated by minimizing the free energy with the constraint that the volume fractions $\phi_i$ should add to one, equivalent to computing the lower convex hull of the free energy surface \cite{wolff_thermodynamic_2011, mao2019phase}.

One can describe nematic ordering using a scalar order parameter, $S$.
Microscopically, $S$ is given by
\begin{equation}
    S = \left\langle \frac{3}{2}\cos^2\theta-\frac{1}{2} \right\rangle,
\end{equation} where $\theta$ is the deviation of a rod-like nematogen from the average orientation, and $\langle \ldots \rangle$ indicates an average over all nematogens.
A value of $S=0$ corresponds to an isotropic liquid (no ordering), and $S=1$ is completely ordered. 
This simple description neglects fluctuations and assumes uniaxial ordering. More complex descriptions are possible (see \cite{wang_critical_1996}, for example). 

The Maier-Saupe model for a pure nematogen fluid describes an energetic preference for orientation competing with an entropic preference for the isotropic state. 
The free energy density is given by 
\begin{equation}\label{eq:MS,pure}
    g = -\frac{1}{2}U S^2 - k_{\mathrm{B}}T\,\sigma(S).
\end{equation}
The parameter $U$ describes a mean-field preference for alignment due to molecular interactions.
The function $\sigma(S)$ describes the change in entropy upon alignment. 
We calculate it numerically from the system of equations in \cite{brochard_phase_1984}.
Details for this calculation are given in Sec.  \ref{sec:appendix_MS_calc} in the appendix.
The equilibrium state is given by the minimum of this function; unlike the composition, we do not need to apply any constraints to $S$.
At high temperatures this function has a minimum at $S=0$, corresponding to an isotropic phase. At the nematic transition ($ 4.54 \,k_{\mathrm{B}} T_{\mathrm{NI}}\approx U$), the minimum jumps to $S\approx 0.429$. 
Plots of the entropy, free energy, and nematic ordering transition are shown in Appendix Fig. \ref{fig:appendix_maier_saupe}

To extend this model to allow for mixtures, we first consider the effects of dilution.
 Since the enthalpy of alignment describes molecular interactions, we assume it scales as $\phi_i^2$. Since the change in entropy of alignment depends on the angular distribution of individual molecules, we assume it scales as $\phi_i$ \cite{shen_spinodals_1995}.
This yields the following free energy for a nematic component $i$ in a mixture:
\begin{equation}
    g_{\mathrm{MS}} 
    = \sum_i-\frac{1}{2} U_i\phi_i^2 S_i^2 - k_BT\, \phi_i \sigma(S_i) .
\end{equation}
As a result, the nematic transition temperature for pure species $i$ is proportional to the volume fraction, \textit{i.e.} $T_{NI,i}=T_{NI}\phi_i$ \cite{reyes_isotropicisotropic_2019, shen_spinodals_1995}.

Finally, we consider the orientation-dependent interactions that occur between  species. 
The generalization of the Maier-Saupe model is done in \cite{brochard_phase_1984} and \cite{palffy-muhoray_mean_1985}, and yields an additional term in the form
\begin{equation}\label{eq:MS,int}
    g_{\mathrm{MS, int}}=\sum_{i<j}\alpha_{ij}\phi_i\phi_jS_iS_j.
\end{equation}
This term reflects the ability of species $j$ to help orient species $i$, for $\alpha_{ij}<0$. A nice interpretation for binary mixtures of nematogens is described in \cite{chiu_phase_1996}. 

Combining the results from the previous paragraphs, the total free energy density for an $N$-component system is
\begin{multline}\label{eq:free_energy}
    g(\{S_i\}, \{\phi_i\}, T) = \\ \sum_i \phi_i\left(\frac{k_{\mathrm{B}}T}{v_i}\ln\phi_i  - k_BT\,\sigma(S_i) -\frac{1}{2}U_i\phi_i S_i^2 \right) + \\
     \sum_{i<j} \phi_i\phi_j\left( A_{ij} + B_{ij}\,T\,  + \alpha_{ij} S_iS_j \right) + 
     \sum_{i< j \le k}\phi_i\phi_j\phi_kC_{ijk} .
\end{multline}

We find the equilibrium state by minimizing the free energy as a function of the free variables (${\phi_i,S_i}$) with respect to constraints ($\sum_i \phi_i = 1$).
Here, we briefly outline the numerical procedure.
First, we define a grid of compositions $\{\phi_i\}$ and temperatures $T$.
For each composition and temperature, we
minimize the function over the non-conserved variables to arrive at an effective free energy of only the fields and conserved variables given by  
\begin{equation}\label{eq:effective_free_energy}
    g_{\mathrm{eff}}(\{\phi_i\}, T)=\text{min}_{\{S_i\}}\,g(\{S_i\}, \{\phi_i\}, T),
\end{equation}
as was done in \cite{garbes_putzel_phenomenological_2008, wolff_thermodynamic_2011} (and implicitly in \cite{riccardi_instability_1998}). 
Note that this process also yields the nematic ordering of each species in the mixture, \textit{i.e.} the fields $\{S_{\mathrm{min},i}(\{\phi_i\},T)\}$.
We calculate the final phase diagrams by convexifying this effective free energy for each temperature.
Regions of demixing are identified by finding simplices that are stretched relative to the initial grid, following the procedure described by \cite{wolff_thermodynamic_2011, mao2019phase}.

\subsection{Comparison of theory and experiment for binary systems}\label{sec:binary_theory}

\begin{figure}[t]
\includegraphics[width=\columnwidth]{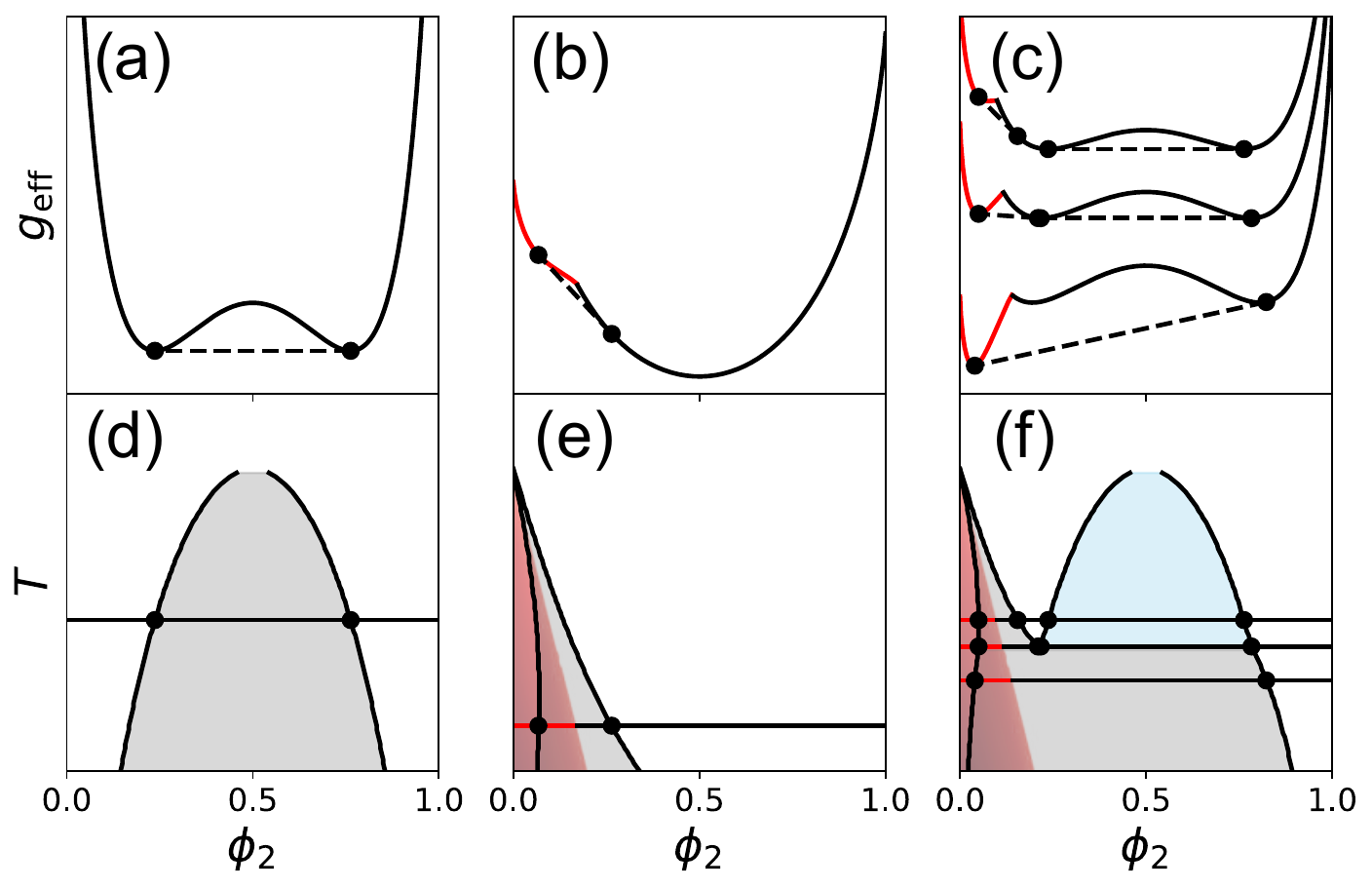}
\caption{\label{fig:minFE} \textbf{Free energy surfaces for nematogen-solvent mixtures have multiple non-convexities leading to phase separation.} (a) Free energy surface and double-tangent construction (\textit{i.e.} convexification) for a binary solvent-solvent mixture with $A_{12}>2$ at a phase-separating temperature  and (d) its corresponding phase diagram. (b) Free energy surface of a binary nematogen-solvent mixture with weak isotropic interactions ($A_{12}=1$, $T_{\mathrm{NI},1}=1$) featuring a non-convexity from the nematic transition, resulting in (e) a phase diagram with isotropic-nematic coexistence. (c) Free energy surface of nematogen-solvent mixture with relatively strong isotropic interactions ($A_{12}=2$, $T_{\mathrm{NI},1}=1$) featuring two phase-separating regions (top), three co-existing phases (middle), or one phase-separating region (bottom) at different temperatures. (f) Phase diagram generated from the free energy surfaces in (c). Shading colors in panels (d), (e), and (f) match Fig. \ref{fig:binary_data}, and the red background color indicates nematic ordering.}
\end{figure}

We illustrate the application of this model with three simple
binary systems,  shown in Figure \ref{fig:minFE}.
In this section, 
we set $B_{12},C_{122},\alpha_{12}=0$ and $v_1,v_2=1$ for simplicity.

We first computed a phase diagram for a phase-separating mixture of simple liquids ($U_1,U_2=0$), shown in Fig. \ref{fig:minFE}a,d.
This behavior can be captured by setting $A_{12}=2$ (\textit{i.e.} standard Flory-Huggins).
The non-convexity in the free energy surface, and hence phase separation, comes from an increase in the free energy due to the unfavorable isotropic solvent interactions described by $A_{ij}>0$.

Results for typical mixture of a nematogen ($U_1=4.54$, so that $T_{NI,1}=1$) with a ``good solvent" ($U_2=0$, $A_{12}=1$) are shown in Fig. \ref{fig:minFE}b,e.  
The non-convexity occurs between the two phases, due to a change in the slope of the free energy surface between the nematic (red) and isotropic (black) portions.
No phase separation occurs between two isotropic phases, even as the temperature approaches zero: any non-convexity in the free energy surface due to isotropic interactions is ``hidden'' by the isotropic-nematic phase separation.
Qualitatively, this reproduces  the features of the 5CB-squalene phase diagram.
Roughly, this behavior occurs whenever the top of the isotropic-isotropic phase separating region (that would occur without any nematic transition) is below the line $T=T_{NI,1}\phi_2$.
Even when the isotropic-isotropic coexistence is hidden, the values of $A_{12}$ and the other isotropic interaction parameters affect the size and shape of the isotropic-nematic coexistence region. 
For example, we used $A_{12}=1$ in panels \ref{fig:minFE}b,e to make the non-convexity in the free energy surface more visible and the phase-separating region larger.

Finally, we calculated the free energy and its derived phase diagram for a nematogen ($U_1=4.54$) and a ``bad solvent" ($U_2=0$, $A_{12}=2$), shown in panels \ref{fig:minFE}c,f.
Above the triple point, having both a strong isotropic interaction ($A_{12}$) and a nematic transition can result in two separate non-convexities in the free energy (top curve in Fig. \ref{fig:minFE}c).
One is caused primarily by the nematic transition, and the other by the isotropic interactions.
This results in  two distinct regions of phase-separation.
At the triple point temperature, where three phases can coexist, a single tangent connects three points on the surface (the middle curve in Fig. \ref{fig:minFE}c).
Below the triple point, there is a single region of demixing between an isotropic and nematic phase.
This system reproduces the qualitative behavior of 5CB-squalane, and this occurs whenever the isotropic interactions are sufficiently large.

These results show that phase separation can appear from two distinct mechanisms.
Phase separation can be driven by  isotropic interactions as in typical liquid-liquid phase separation.
Alternatively,  alignment  increases the effective attractive between nematogens.
At low temperatures (\textit{e.g.} below the triple point in panels \ref{fig:minFE}c,f), these two effects combine and form a larger difference in equilibrium compositions than expected from either effect on its own.

By varying the strength of the isotropic interactions, our model can easily recreate the qualitative features of phase diagrams we observe in our experiments. 
However, to accurately describe the real phase boundaries extra phenomenological terms are needed. 
The Flory-Huggins parameters ($A_{ij}, B_{ij}, C_{ijk}$) generally control the shape of the isotropic-isotropic phase boundary \cite{fernandez2026thermodynamics}, while the Maier-Saupe interaction parameters ($\alpha_{ij}$) control the slope and curvature of the isotropic-nematic coexistence region extending from the pure nematogen nematic transition (see \cite{chiu_phase_1996} for more details; the effect here is similar to their modeling of binary nematogen mixtures). 
Adjusting these to fit experimental data is relatively straightforward due the distinct effects of each parameter.

After adjusting these parameters by hand, the mean-field model accurately fits the 5CB-squalene and 5CB-squalane phase diagrams, as shown in Fig. \ref{fig:binary_data}a,b. 
The fit of data to the theory proceeded as follows.
The nematic transition temperature of 5CB was set to $k_BT=1$ (corresponding to $U=4.54$, from eq. \ref{eq:MS,pure}) for convenience during the calculations, and rescaled to 35$^\circ$ C to match the data. 
The molar volume of 5CB was set to 1, yielding volumes of $2.1$ and $1.9$ for squalane and squalene, respectively, based on their nominal densities at room temperature (see Methods).
The remaining parameters were adjusted by eye, where the indices $1,2,3$ refer to 5CB, squalene, and squalane, respectively: $A_{12}=1.9, B_{12}=-1.2, C_{122}=0.4,A_{13}=6.7, B_{13}=-5.3,$, $C_{133}=0.5$, $ \alpha_{12}=-2.6$, and $\alpha_{13}=-2.3$.
We found that changing $U_2$ and $U_3$ had almost no effect on the phase boundary, unless it was comparable to $U_1=4.54$ (the value for the nematogen), so they were left as 0.
The width of the isotropic-nematic region was sensitive to the relative molar volumes $v_2$ and $v_3$, and the fit in that region could be improved by varying those parameters (See Appendix Fig. \ref{fig:appendix_fit_varying_mol_vols}).

One significant feature of the fitting was that the solvent required  significant nematic interaction with the nematogen in order for the slope of the isotropic-nematic demixing region to be match experiments, \textit{i.e.} it required $\alpha_{12},\alpha_{13}<0$.
If no interaction is included, the nematic transition occurs at $T_{\mathrm{NI}}(\phi_{\mathrm{solvent}})=T_{\mathrm{NI}}(0)\phi_{\mathrm{solvent}}$.
This predicts the nematic phase to appear at much lower temperatures than observed (See Appendix Fig. \ref{fig:appendix_fit_no_alpha}).
Naively, this suggests both that squalane and squalene are aligned in the nematic phase and they have some preference towards alignment. 
Our fit implies values of $S_2$ and $S_3$ (ordering for the non-nematogen solvent) to be around half of the value of $S_1$ (ordering for the nematogen) in the nematic phase.
Quantitative values at one temperature are shown in Appendix Fig. \ref{fig:appendix_solvent_ordering}.

\subsection{Comparison of theory and experiment for ternary systems}

We can now extend our calculations to ternary systems, to both interpret our experimental results and test our model's validity for more complex mixtures. 
Our model and numerical methods are generalizable to any
number of component, so this procedure was difficult only
due to the scaling of the number of points needed to
sample the space (growing exponentially in the number
of species). 
For three species, the phase diagram was still easily computable.

As the number of species increases, the number of parameters increases and their effects on the phase behavior becomes more complex.
However, fits of the binary mixture fix many of the interaction parameters. 
We first set the nematogen-solvent interaction parameters (\emph{e.g.} $A_{12}$, $A_{13}$) to match the fitted values from our binary experiments (see section \ref{sec:binary_theory}).
This leaves use with a set of 5 unknown parameters ($A_{23}, B_{23}, C_{233}, C_{123}, \alpha_{23}$).
The Maier-Saupe interaction between solvents ($\alpha_{23}$) had almost no effect on the phase boundary, so this was kept as zero.
The binary squalane-squalene interaction parameters could not directly measured, apart from the constraint that the mixtures are miscible down to $-80^\circ$ C.
For simplicity, $B_{23}$ and $C_{233}$ were both fixed at zero.
We were able to move the computed phase boundaries to fit our experimental data by setting the remaining two parameters to $A_{23}=0.4$ and $C_{123}=-1.1$.
(The computed phase diagram without these interaction parameters is plotted in Appendix Fig. \ref{fig:appendix_ternary_fit_no_extra_params}, showing significantly worse agreement with measurements.)

\begin{figure*}[t]
\includegraphics[width=0.7\textwidth]{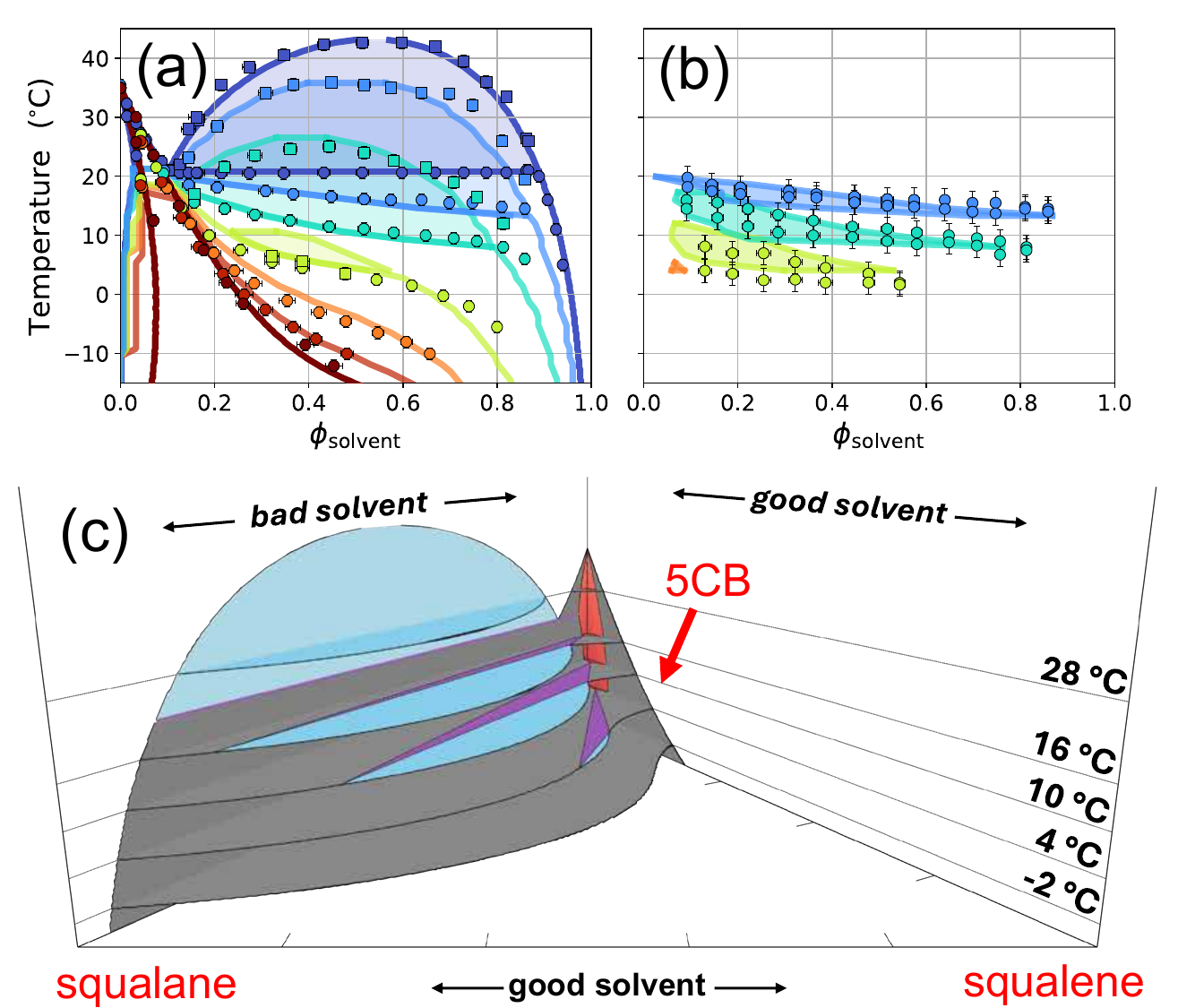}
\caption{\label{fig:3D_ternary} 
\textbf{Fit of model to ternary 5CB-squalane-squalene mixtures.}
%
(a) Comparison of the measured regions of phase separation (points) to the fit described in the text (smooth curves), differentiating between regions of purely isotropic liquids (square points, shaded region) and regions which contain a nematic phase (circles, non-shaded). 
(b) Comparison of the measured regions of three-phase coexistence (points) with the predicted regions (shaded region enclosed by curves). In (a) and (b), each color corresponds to a different solvent mixture, from pure squalane (dark blue) to pure squalene (dark red). The compositions are the same as in Fig. \ref{fig:ternary_expt}.
(c) Fitted phase diagram, displayed in three dimensions.
The binary phase diagrams computed in Fig. \ref{fig:binary_data} are plotted on the 5CB-squalane and 5CB-squalene boundaries.
Isotropic-isotropic coexistence regions is indicated in gray, the isotropic-nematic in blue, and the three-phase region in purple.
}
\end{figure*}

The phase boundaries produced by our model are able to reproduce the experimental results for the solvent mixtures, as shown in Fig. \ref{fig:3D_ternary}a,b. 
The smooth curves are computed phase boundaries and points are from experiments (also shown in Fig. \ref{fig:ternary_expt}).
In Fig. \ref{fig:3D_ternary}a, we compare the regions where the mixture phase separates for the model and experiments, differentiating between when the mixture is fully isotropic (shaded region) or contains a nematic phase (non-shaded region).
Square points indicate isotropic-isotropic phase separation and circular points indicate the appearance of a nematic phase (on cooling).
The model reproduces both the shrinking isotropic-isotropic coexistence region as the amount of squalene increases, and the location of the appearance of the nematic phase.
Fig. \ref{fig:3D_ternary}b compares experimental and computed boundaries for three-phase coexistence, which also show good agreement.

The effect of using a solvent mixture can be seen more clearly by examining the full phase diagram in three dimensions. 
Ternary phase diagrams at various temperatures are plotted in Fig. \ref{fig:3D_ternary}c, aligned with the binary phase diagrams (from Fig. \ref{fig:binary_data}).
As in Fig. \ref{fig:binary_data}, the blue region corresponds to isotropic-isotropic coexistence, the gray region to isotropic-nematic coexistence, and the purple region to three-phase isotropic-isotropic-nematic coexistence.
Above the triple point temperature for the 5CB-squalane mixture ($>21^\circ$ C), there is a region of isotropic-isotropic coexistence and a disconnected region of isotropic-nematic coexistence.
At the triple point temperature ($\approx 21^\circ$ C), these regions meet at the 5CB-squalane triple point. 
Cooling further, the triple point moves towards the ``interior'' of the phase diagram, becoming a region of three-phase coexistence.
Each edge of the three-phase region connects to a two-phase region: one edge is an isotropic-isotropic coexistence region, two are isotropic-nematic.
The three-phase region and the isotropic-isotropic region both vanish as the temperature is lowered, leaving only a single region of isotropic-nematic coexistence. 

Quantitatively, the ternary mixture is modeled quite well using knowledge of the binary mixtures. 
The addition of only two extra parameters ($A_{23}, C_{123}$) was able to fit the five additional pseudo-binary phase diagrams measured with solvent mixtures.
While the ``volume'' of phase space scales exponentially in the number of species, this suggests that the increase in the number of species may not require a large increase in the complexity of the system.


\subsection{Calculations of complex, experimentally accessible phase diagrams}

Having validated our model for at least one experimental system, we surveyed the range of accessible phase diagrams that one might expect. 
With 14 (at least) free parameters, the space of potential phase diagrams for ternary nematogen-solvent-solvent systems predicted from our model is huge.  
As an initial exploration of this high-dimensional space, we decided to focus on the effect of the three pairs of binary interactions ($A_{12},A_{23},A_{13}$) for a single nematic species ($U_1=4.54$).
This is inspired by our classification of binary mixtures as a nematogen mixed with either a ``good solvent'' or a ``bad solvent,'' which can be controlled by varying $A_{ij}$ (small for a ``good solvent'', large for a ``bad solvent'').
The remaining interaction parameters ($B_{ij}, C_{ijk}, \alpha_{ij}, U_2,U_3$) were set to zero and the molecular volumes ($v_i$) were kept at one.

Categorizing each of the three pairs of interactions as ``good'' or ``bad'' yields six possible ternary mixtures.
For each of these six sets of interactions, we chose values of $A_{12},A_{23}$ and $A_{13}$ to get a binary phase diagram with the correct solvent quality.
We avoided setting two values of $A_{ij}$ to the same number (\textit{e.g.} avoiding $A_{12}=A_{13}$).

\begin{figure}[ht!]
\includegraphics[width=\columnwidth]{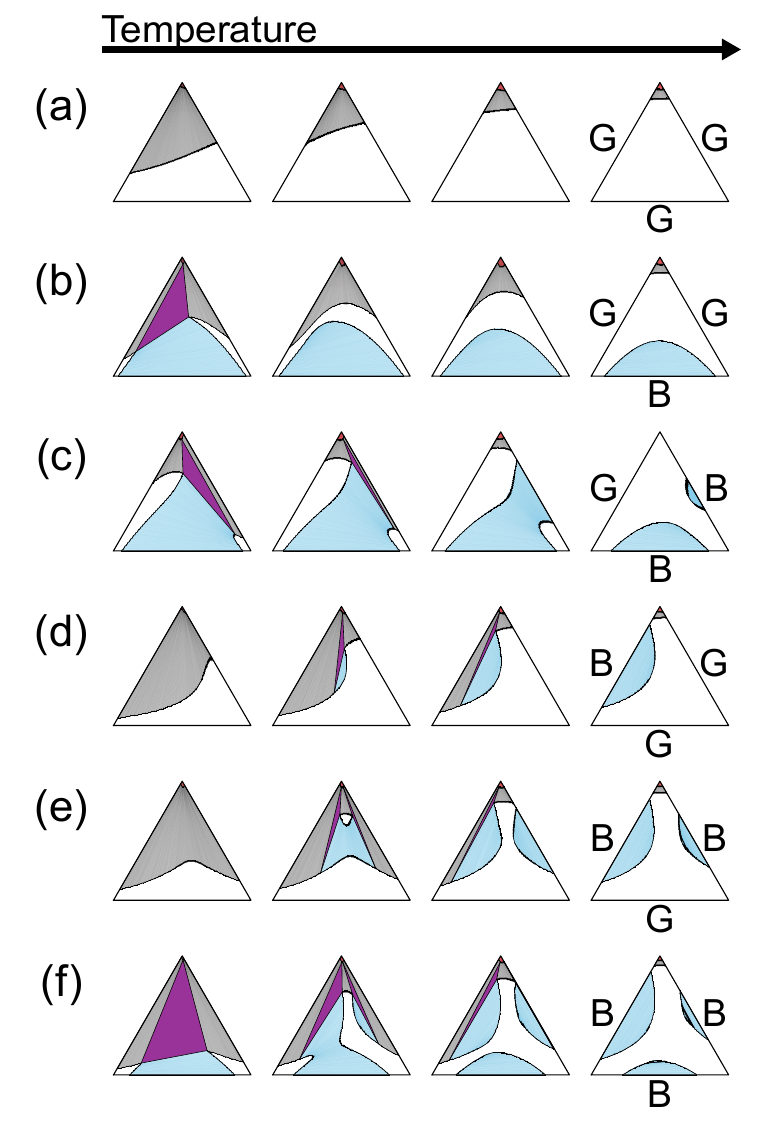}
\caption{\label{fig:solvent_quality}
\textbf{Diverse phase diagram structures can be realized from nematogen-solvent-solvent mixtures.}
(a-f) Ternary (nematogen-solvent-solvent) phase diagrams with varying pairwise interactions.
The nematogen (index 1) is the top corner, and the left (2) and right (3) corners correspond to the two solvents.
The edges of the rightmost phase diagram are labeled based on the quality of the pairwise interactions: ``G'' for ``good'' (small $A_{ij}$), ``B'' for ``bad'' (large $A_{ij}$).
For these examples, the pairwise interactions $(A_{12}, A_{13}, A_{23})$ and temperatures are as follows:
(a) (1.4, 1.0, 0.0) plotted at $T=0.75, 0.8, 0.85, 0.9$.
(b) (1.4, 1.0, 2.5) plotted at $T=0.7, 0.75, 0.8, 0.9$.
(c) (1.3, 2.0, 2.5) plotted at $T=0.8, 0.84, 0.88, 0.93$.
(d) (2.3, 1.3, 0.5) plotted at $T=0.8, 0.85, 0.9, 1.0$.
(e) (2.3, 2.0, 1.5) plotted at $T=0.85, 0.86, 0.89, 0.93$.
(f) (2.4, 2.0, 2.1) plotted at $T=0.8, 0.85, 0.88, 0.94$.
}
\end{figure}

Constant-temperature slices of the six resulting phase diagrams are shown in Fig. \ref{fig:solvent_quality}.
Within each panel, the temperature is increased  from left to right.
Temperatures were chosen to best reflect the change in the phase diagram topologies.
Regions within the phase diagram are colored based on the color scheme used in previous figures (\textit{e.g. } Fig. \ref{fig:3D_ternary}a).
Fig. \ref{fig:solvent_quality}d corresponds (qualitatively) to the 5CB-squalane-squalene system.

The simplest phase diagram is with three good solvent interactions, shown in Fig. \ref{fig:solvent_quality}a. 
There is only one region of isotropic-nematic phase separation, which starts from the nematic transition of the nematogen and extends into the interior of the phase diagram as the temperature is lowered.
For this mixture, it seems possible to fit the phase boundaries of various solvent mixtures using the binary model, since there are no three-component features that appear.

The remaining phase diagrams (Fig. \ref{fig:solvent_quality}b-f) all have three-phase coexistence that appears at the intersection of isotropic-isotropic and isotropic-nematic coexistence.
Generally, the three-phase region starts from a triple point in one of the binary mixtures.
In one case (Fig. \ref{fig:solvent_quality}b) the three-phase region appears as the isotropic-isotropic critical point intersects the isotropic-nematic coexistence region.
The existence of three-phase coexistence at low temperatures depends on the interaction between the two solvents; solvent pairs with a ``good'' interaction revert to isotropic-nematic coexistence after cooling (Fig. \ref{fig:solvent_quality}a,d,e), while those with a ``bad'' interaction maintain a three-phase region even as the temperature drops (Fig. \ref{fig:solvent_quality}b,c,f).
Two of the phase diagrams have two binary-mixture triple points (Fig. \ref{fig:solvent_quality}e and \ref{fig:solvent_quality}f), resulting in two three-phase regions.
In Fig. \ref{fig:solvent_quality}e, the regions vanish after intersecting, while one three-phase region persists at lower temperatures in Fig. \ref{fig:solvent_quality}f.

\section{Conclusion}

This work broadly investigates the phase behavior of nematogen-solvent mixtures.
Binary nematogen-solvent mixtures can generally be classified as either a nematogen in a bad solvent or a good solvent, \textit{i.e.} whether the phase diagram has a region of isotropic-isotropic coexistence or not.
Phase boundaries can be continuously tuned between the two kinds of phase diagrams by using a solvent blend.
However, this results in a three-phase region, connected to the triple point in the binary nematogen-bad solvent phase diagram.
The behavior of binary and ternary mixtures can be quantitatively described by combining a phenomenological Flory-Huggins model of isotropic liquids with the Maier-Saupe model of nematic ordering.
Varying solvent parameters, a wide variety of ternary phase diagrams are predicted.

These findings have a number of implications.
First,  fitting of measured phase diagrams suggests that non-nematogen species  have some ``latent'' tendency to align, resulting in the co-alignment of solvents in the nematic phase.
Early work on liquid crystals supports this hypothesis \cite{dave_mixed_1954, walter_zur_1925} and future work could investigate this directly at a microscopic level.
Second,  solvent blends offer a simplified pathway to target specific domain morphologies  \cite{lapena_effect_1999} without discrete change chemical changes \cite{browne_structural_2025}.
Finally, coupling an additional phase transition to fluid phase separation provides an approach for creating phase diagrams with complex topologies.
As hypothesized in \cite{murugan2026information}, a large number of phase boundaries could make these phase diagrams useful for information processing.

\section{Author contributions}
SBD and ERD conceptualized the project, designed the methodology, interpreted the results, wrote the manuscript, and visualized the data. TM and SBD developed the software to compute phase diagrams. All experiments, data analysis, and computations were performed by SBD with input from ERD. 

\section{Conflicts of interest}
The authors do not declare any conflict of interest.

\section{Data availability}
Data and code used for this work will be available on Zenodo. 
The code will also be available on Github.

\section{Acknowledgements}

We thank Kaarthik Varma, James Sethna, Fran\c{c}oise Brochard-Wyart, and Rob Style for helpful discussions about this work, specifically on the thermodynamic modeling. 
We would also like to thank Youlim Ha, Nick Abbott, and Chinedum Osuji, who all provided valuable experimental guidance (particularly in the choice of squalane as a solvent).
All members of the Dufresne lab provided general advice and feedback on experimental methods and results.
ChatGPT and Claude were used in a few instances to generate Python code.

\section{Methods}

\subsection{Sample preparation}

All samples were made from varying amounts of 4’-pentyl-4-cyanobiphenyl (5CB) (97\%, Astatech), squalane ($>98$\%, TCI), and squalene ($>98$\%, TCI).
For phase diagram measurements, sets of samples were prepared \textit{via} serial dilution.
Samples compositions were determined by mass, and the volume fraction using the following values for the nominal densities at room temperature of 5CB, squalene, and squalane, respectively: 1.01 g/mL \cite{deschamps_vapor_2008}, 0.86 g/mL \cite{mylona_reference_2014}, 0.81 g/mL \cite{popa_methods_2015}.
When removing any liquid from a mixture for a measurement, the sample was kept at sufficient temperature to prevent phase separation.

\subsection{Temperature-controlled polarized optical microscopy}

Phase boundaries were determined using temperature-controlled polarized optical microscopy (POM) measurements of samples in glass capillaries.
The polarizer and analyzer were set to be near perpendicular (usually at $70^\circ$ to each other), so that nematic phases were visibly brighter than isotropic phases. 
Images were captured with a Hamamatsu CMOS camera (ORCA Flash 4.0 v2, C11440-22CU).
Phase diagram measurements were made using a 2x air objective (Nikon, Plan Apo) and images of phase coexistence were captured with a 20x air objective (Nikon, Plan Fluor, ELWD).

Samples were loaded into glass capillaries from Electron Microscopy Sciences (Cat. \#63823-10), placed onto a custom aluminum capillary holder, and the temperature was controlled using an INSTEC heating stage (TSA12Gi).
The aluminum holder was used to minimize temperature gradients across the optical path.
Compositional gradients appeared once the mixture phase separates, and persisted even when heated to a temperature where the mixture is one-phase.
To keep samples homogeneous, samples were kept warm enough to prevent phase-separated while they were loaded into capillaries and the microscope and measurements were done while cooling.

During measurements of phase boundaries, samples were were cooled slowly ($\le 0.5^\circ$ C/min) and held at fixed temperature for at least one minute at each half degree to allow equilibration. Error bars for temperature are (over)estimated based on these step sizes as $\pm 1 ^\circ$ C.

%

\clearpage

\section{Appendix}

\setcounter{figure}{0}

\subsection{Stability of three-phase coexistence}

\begin{figure}[!htbp]
\includegraphics[width=\columnwidth]{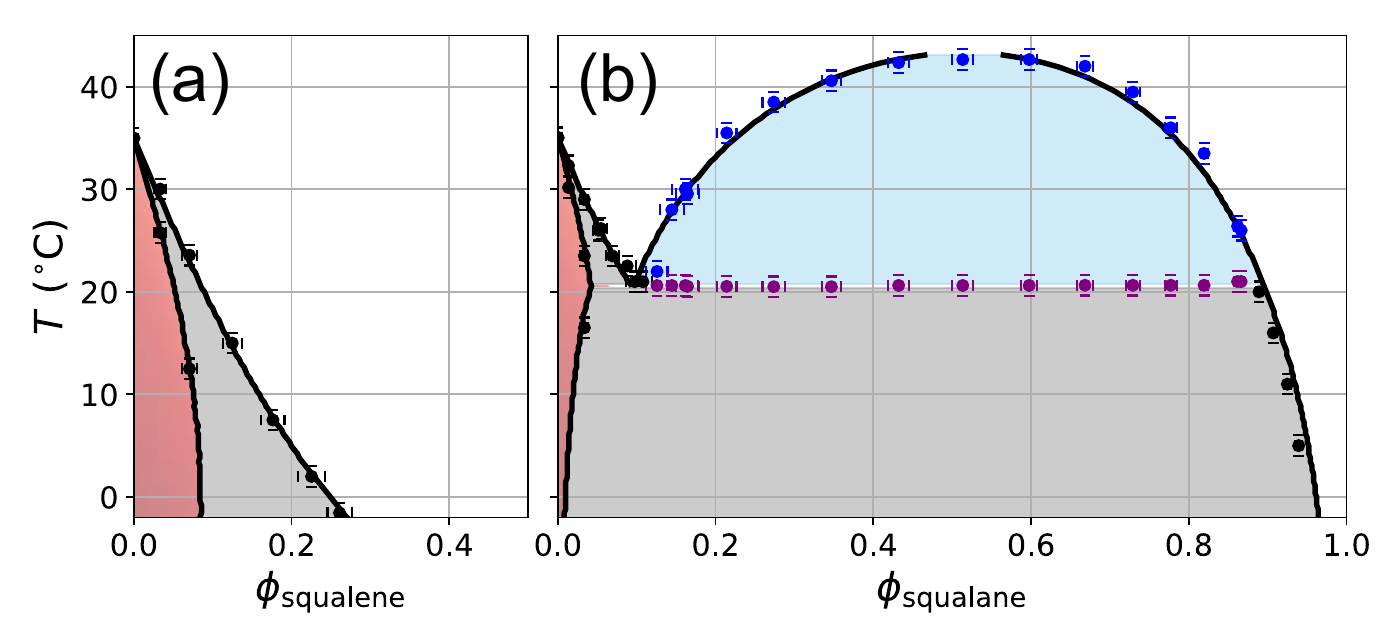}
\caption{\label{fig:appendix_three_phase_overnight} Three phases coexist for multiple hours at constant temperature. Images of a sample with $\phi_{5CB}=0.75$, $\phi_{\mathrm{squalane}}=\phi_{\mathrm{squalene}}\approx0.125$ are taken with partially crossed polarizers. The was cooled to $5.5^\circ$ (a), then held at the same temperature for 2 hours (b) and 12 hours (c).
While the phases coarsen and drift slightly over time time, three distinct phases are visible for the entire time period.
} 
\end{figure}

\subsection{Maier-Saupe calculation of the alignment entropy}\label{sec:appendix_MS_calc}

Here, we briefly outline the method used to calculate the change in entropy due to alignment from the Maier-Saupe model, as described in \cite{brochard_phase_1984}.

The orientation of each rod-like nematogen is described by the angle $\theta$ from the director, and follows a trial distribution function $f(\theta)$ given by 
\[
    f(\theta) = \frac{1}{Z}\text{exp}\left(m\frac{3\cos^2\theta-1}{2}\right),
\]
where the normalization $Z$ is given by 
\begin{equation}\label{eq:Z_m}
    Z(m) = \int d\Omega\; f(\theta)
\end{equation}
The variable $Z$ is a normalization constant and the parameter $m$ describes the degree of alignment. 
The nematic order parameter $S$ is given by 
\begin{equation}\label{eq:S_m}
    S(m) = \int d\Omega\; f(\theta) \frac{3\cos^2\theta-1}{2}.
\end{equation}

To find the equilibrium values of $m$, and hence $S$, we want to calculate the energy and entropy of alignment, relative to the unaligned, reference state. 
The entropy change due to alignment $\sigma$ is calculated as 
\begin{multline}\label{eq:sigma_MS}
    \sigma = -k_B\int d\Omega\; f \ln f - k_B\int d\Omega\; \frac{1}{4\pi}\ln \left(\frac{1}{4\pi}\right) \\
    = -k_B\int d\Omega\; f \ln (4\pi f)
    = -k\left( \ln\left(\frac{4\pi}{Z}\right) + m\,S\right).
\end{multline}

This system has no analytical solution, but the entropy can be calculated numerically for each choice of $m$.
The values of $Z(m)$ and $S(m)$ can be computed by numerically integrating eqs. \ref{eq:Z_m} and \ref{eq:S_m}, and the entropy calculated \textit{via} eq. \ref{eq:sigma_MS}.
After computing $\sigma$ and $S$ for a range of $m$ values, the function $\sigma(S)$ is computed using a linear interpolation from the pre-computed points.

\subsection{Plots of Maier-Saupe entropy and free energy}

\begin{figure}[!htbp]
\includegraphics[width=\columnwidth]{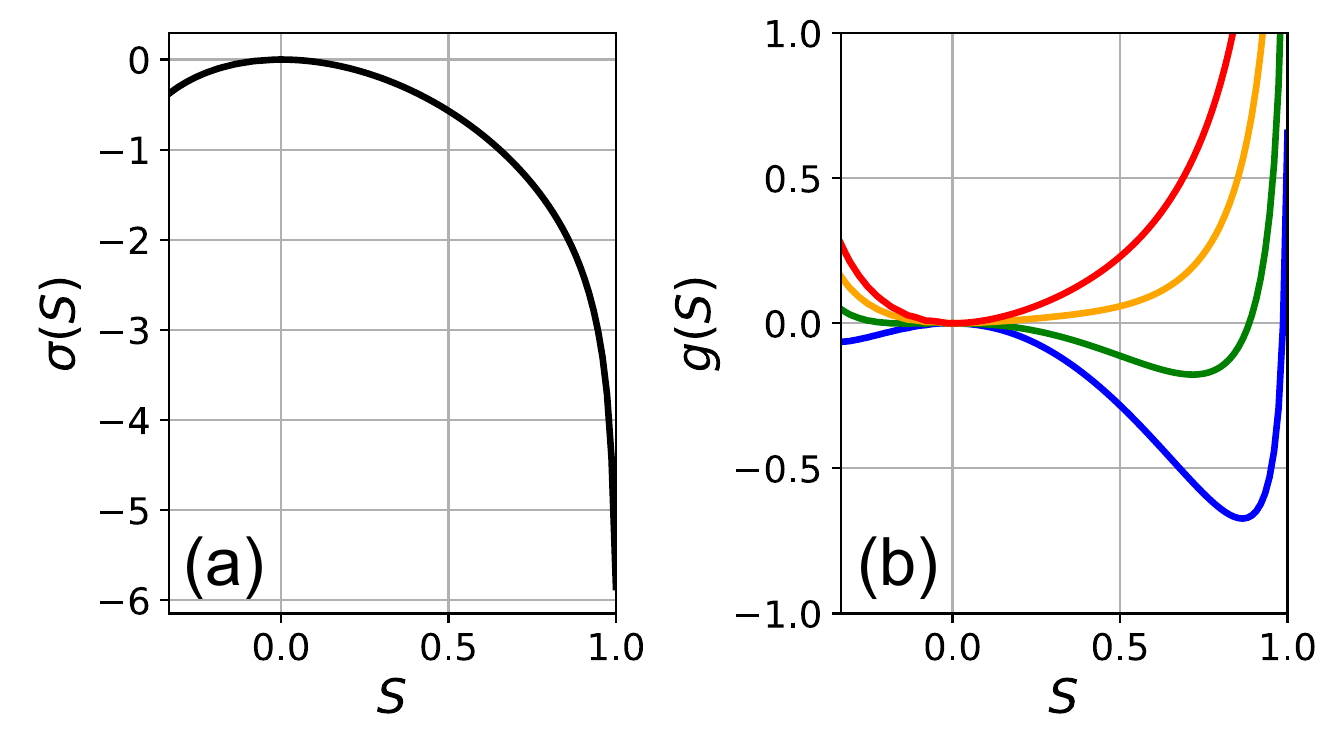}
\caption{\label{fig:appendix_maier_saupe}
Plots of (a) $\sigma(S)$, the entropy change due to alignment, and (b) $g(S)$, change in free energy at varying temperatures due to alignment, as calculated from the Maier-Saupe model. The curves in panel (b) were computed with eq. \ref{eq:MS,pure} using $U=4.54$ so that the nematic transition occurs at $T=1$. Each curve corresponds to a different temperature: $T=1.4$ (red), 1.1 (orange), 0.8 (green), and 0.5 (blue). 
}
\end{figure}

\subsection{Fitting with varied molecular volumes}

\begin{figure}[!htbp]
\includegraphics[width=\columnwidth]{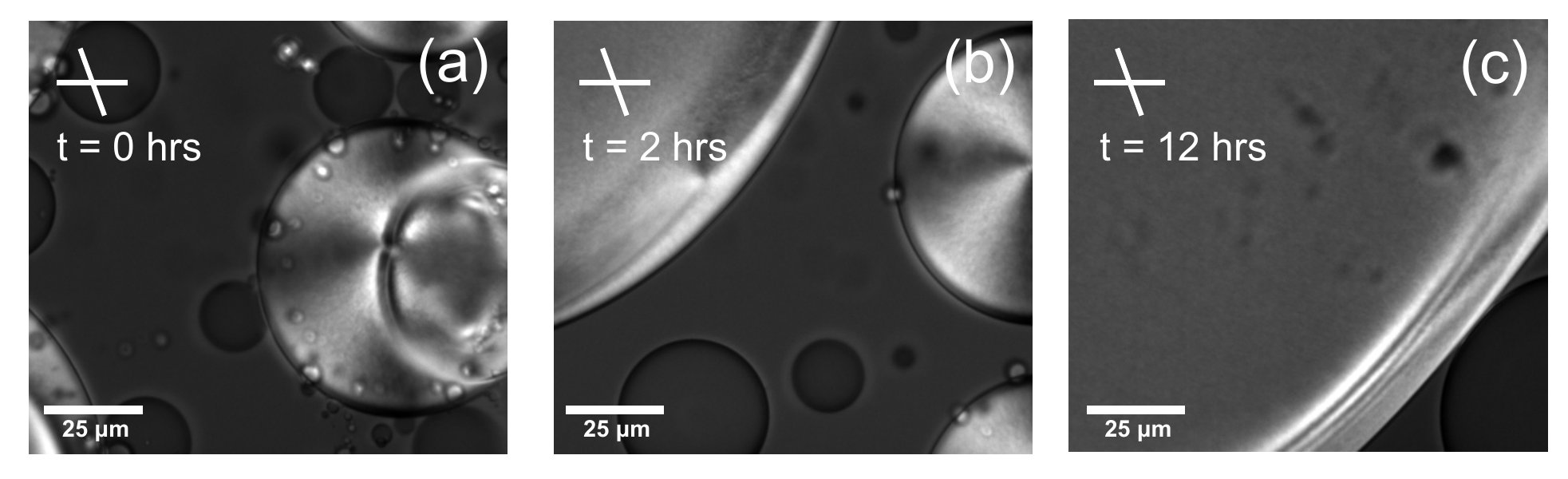}
\caption{\label{fig:appendix_fit_varying_mol_vols} Phase diagrams of (a) 5CB-squalene and (b) 5CB-squalane, allowing the molecular volumes to vary in the fitting. In particular, this is able to improve the fit in the low $\phi_{\mathrm{solvent}}$ region slightly. Lines and regions are from the calculated phase diagram, showing one-phase regions of nematic (red) and isotropic (white), and  of isotropic-isotropic coexistence (gray) and isotropic-nematic coexistence (blue). Data points from measurements are shown as points with error bars. The 5CB-squalene phase diagram was fit with the following parameters: $A_{12}=2.1,B_{12}=-1,C_{122}=-0.2,\alpha_{12}=-2.6$ and $v_1=1.25$. The 5CB-squalane phase diagram was fit with $A_{12}=8.03,B_{12}=-6.3,C_{122}=0.3,\alpha_{12}=-2.5$ and $v_1=1.5$}
\end{figure}

\subsection{Fitting without solvent alignment}

\begin{figure}[!htbp]
\includegraphics[width=\columnwidth]{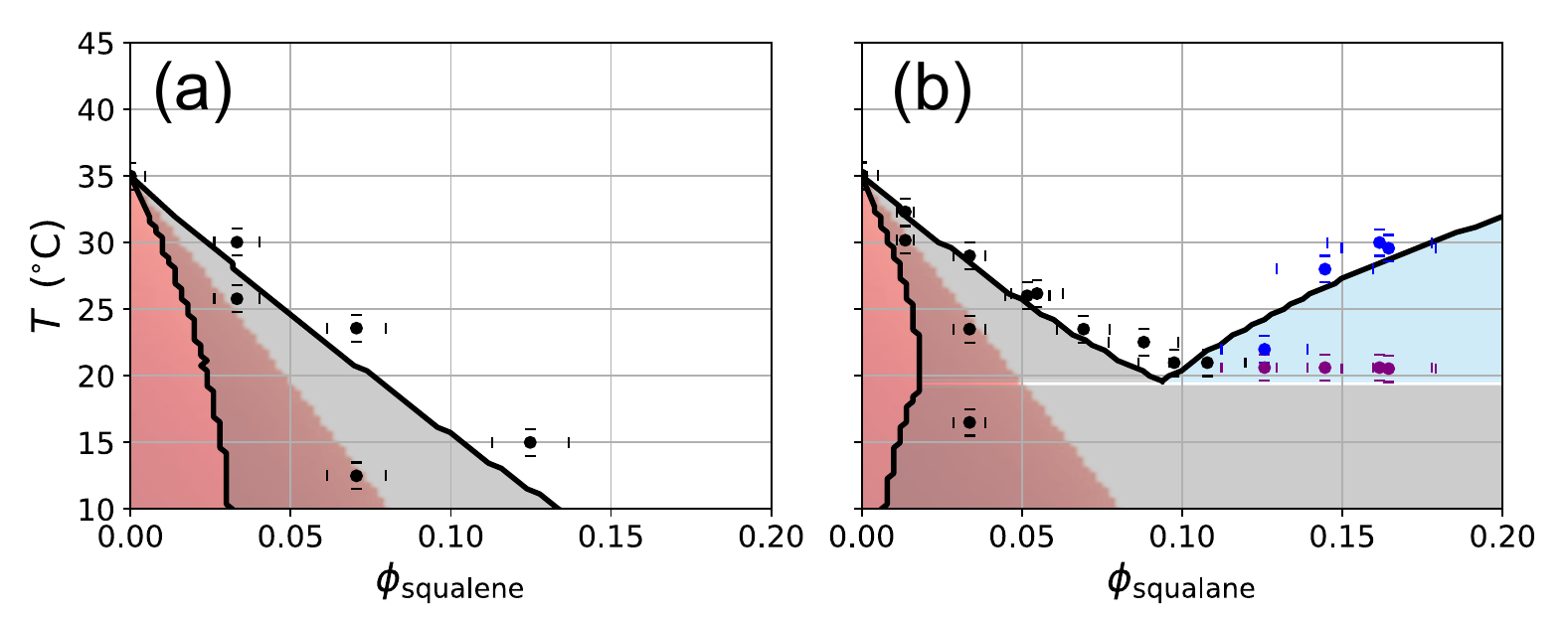}
\caption{\label{fig:appendix_fit_no_alpha}
Fitting phase diagrams without any solvent alignment cannot reproduce the isotropic-nematic phase-coexistence observed in mixtures of 5CB-squalene (a) and 5CB-squalane (b). 
Points are from experiments, and curves and shaded regions are calculated. 
Coloring is identical to Fig. \ref{fig:binary_data}, except that the two-phase region has been made slightly transparent to show the predicted region of nematic ordering (red background).
Parameters are identical to those used in Fig. \ref{fig:binary_data} (described in Sec. \ref{sec:binary_theory}), except with $\alpha_{12}=0$. 
}
\end{figure}

\subsection{Solvent ordering predicted from modeling}

\begin{figure}[!h]
\includegraphics[width=\columnwidth]{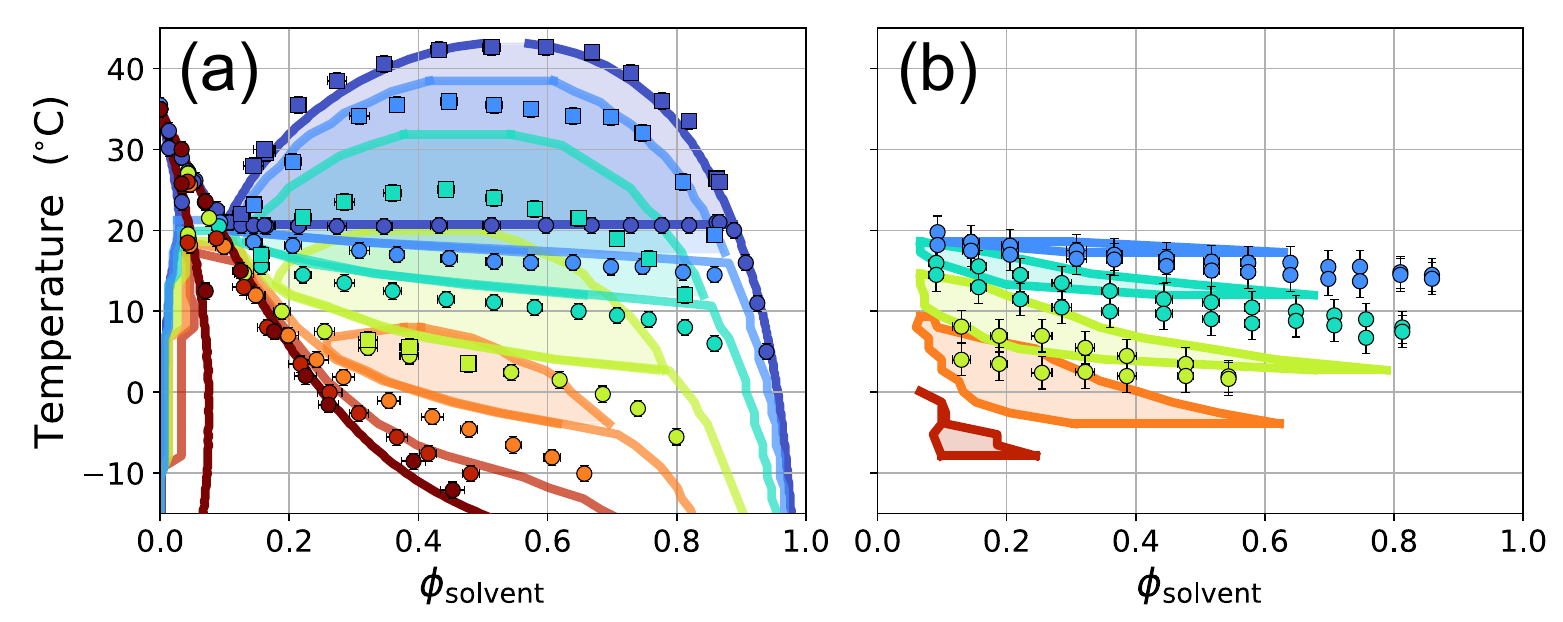}
\caption{\label{fig:appendix_solvent_ordering}
Values of the nematic ordering parameter $S$ computed for the calculated phase diagrams shown in Fig. \ref{fig:binary_data} for 5CB-squalene mixtures (a) and 5CB-squalane mixtures (b) at $T=25^\circ$ C for both the nematogen and solvent.
Note that these values are not all physically accessible; some of these occur within the two-phase region.
}
\end{figure}

\newpage

\subsection{Ternary phase diagram without additional parameters}

\begin{figure}[!h]
\includegraphics[width=\columnwidth]{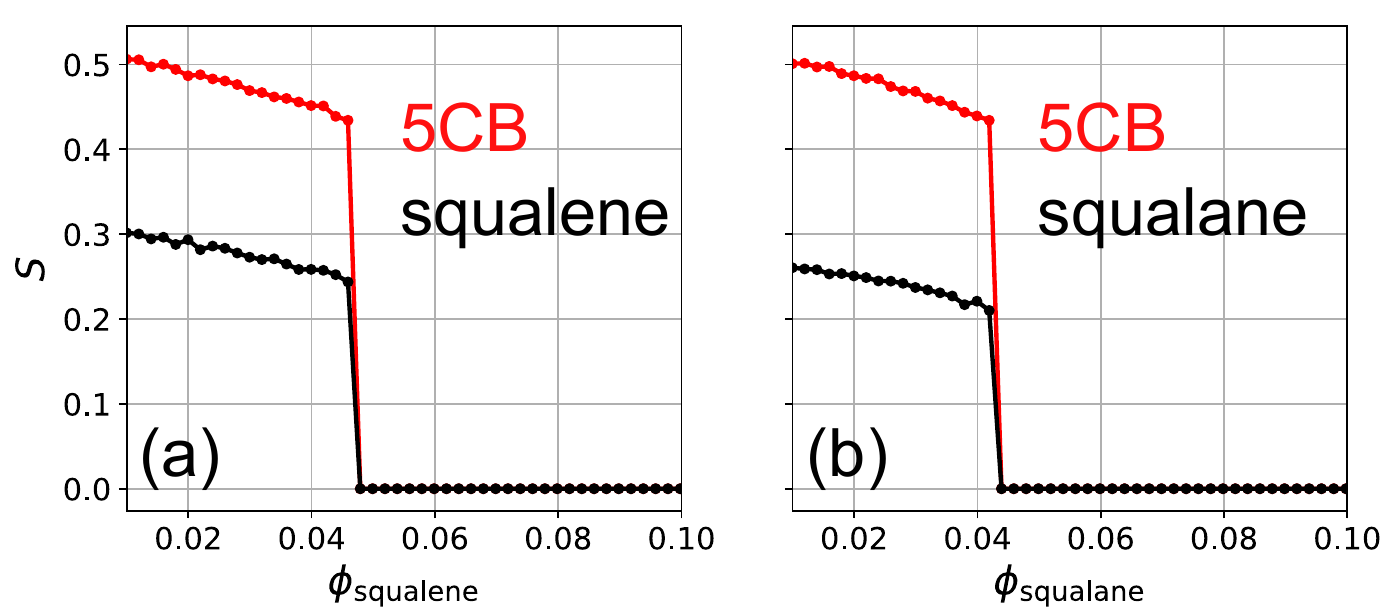}
\caption{\label{fig:appendix_ternary_fit_no_extra_params}
Regions from the computed ternary phase diagram using only parameters fit from the binary mixtures are compared with experimental results.
The coloring, points, lines, and shaded regions have the same meaning as in Fig. \ref{fig:3D_ternary}. 
Fitting parameters are all zero, except for the interaction parameters and molecular volumes described in Sec. \ref{sec:binary_theory}.
}
\end{figure}

\end{document}